# Mirrors for space telescopes: degradation issues

**Denis Garoli[1]\*, Luis V. Rodriguez De Marcos[2], Juan I. Larruquert[3], Alain J. Corso[4], Remo Proietti Zaccaria[5,6], and Maria G. Pelizzo[4]**

1. Faculty of Science and Technology Free University of Bozen, Piazza Università 5, 39100 Bolzano, Italy
2. The Catholic University of America and NASA Goddard Space Flight Center (CRESST II), 620 Michigan Ave., N.E. Washington, DC 20064, USA
3. GOLD-Instituto de Optica-Consejo Superior de Investigaciones Científicas, Serrano 144, Madrid 28006, Spain
4. CNR-IFN, Via Trasea 7, Padova, Italy
5. Istituto Italiano di Tecnologia, Via Morego 30, 16136 Genova, Italy
6. Cixi Institute of Biomedical Engineering, Ningbo Institute of Materials Technology and Engineering, Chinese Academy of Sciences, Zhongguan West Road 1219, Ningbo 315201, China

\* Correspondence: denis.garoli@unibz.it

**Abstract:** Mirrors are a subset of optical components essential for the success of current and future space missions. Most of the telescopes for space programs ranging from Earth Observation to Astrophysics and covering all the electromagnetic spectrum from X-rays to Far-Infrared are based on reflective optics. Mirrors operate in diverse and harsh environments that range from Low-Earth Orbit, to interplanetary orbits and the deep space. The operational life of space observatories spans from minutes (sounding rockets) to decades (large observatories), and the performance of the mirrors within the optical system is susceptible to degrade, which results in a transient optical efficiency of the instrument. The degradation that occurs in space environments depends on the operational life on the orbital properties of the space mission, and it reduces the total system throughput and hence compromises the science return. Therefore, the knowledge of potential degradation physical mechanisms, how they affect mirror performance, and how to prevent it, is of paramount importance to ensure the long-term success of space telescopes. In this review we report an overview on current mirror technology for space missions with a particular focus on the importance of degradation and radiation resistance of the coating materials. Particular detail will be given to degradation effects on mirrors for the far and extreme UV as in these ranges the degradation is enhanced by the strong absorption of most contaminants.

**Keywords:** space optics; mirrors; coatings; radiation; thin film; multilayer; degradation; contamination; optical systems.

---

**1. Mirror Technology**

The trend for the future space missions is the use of high-resolution, large bandwidth telescopes [1][2][3]. This will require new optical systems with large apertures and extreme operation conditions. Examples are mission concepts such as LUVOIR, HabEx, Galaxy Evolution Probe, and the X-Ray Observatories [4][5,6][7]. These and many other present or future space concepts [8] introduce new challenges in mirror technologies, from the optical design, to the substrate and the coatings. Mirrors are critical components in space telescopes, which are extensively used for the observations of Earth and astronomical objects. Mirror technology is evolving continuously due to improvements in materials, design, manufacture and metrology. The main advantages of mirrors with respect to refractive optics such as lenses are the following: (i) they can work over a very wide spectral bandwidth (achromat); (ii) they can be manufactured with different shapes and large dimensions compared to lenses; (iii) they are suitable for scanning devices; (iv) for some applications such as X-ray optics, grazing incidence mirrors are the only option available. Future large telescopes will cover an increased spectral range of observation with a broad range of multi-spectral and hyper-spectral instruments, and this can be achieved only with reflective telescopes.

A mirror consists in a substrate and, most often, a coating. Substrates can be selected among a limited number of materials. Fundamental parameters are: i) Specific stiffness; ii) thermal stability, iii) space environmental resistance, iv) achievable surface quality, v) weight, and vi) financial aspects. Regarding the choice of mirror substrates, extensive work has been performed and reported [9][10][2][11][12].

Al or Al alloys, Be, Si, SiC, Zerodur®, nickel and fused silica have been ~~used~~ employed [10][2][3][12][13], although glass has been the most used material for mirror substrates wing to its thermal stability and ease engineering into high-quality optical surface [14]; for instance, it is used as substrate in the Hubble Space Telescope, the largest space mirror still operating. However, one important shortcoming of using glass is its weight, which often limits its use to small aperture mirrors. For this reason, new materials have been developed with the near future state-of-the-art mirror research focusing on segmented mirrors prepared on Zerodur, Be, Al, Si or SiC substrates [14]. New large telescopes with active mirrors are now developed with carbon based (lightweight) materials. Silicon Carbide (SiC), in particular, has been successfully used in ESA Herschel Space Observatory [15] and it's still extensively investigated as potential standard because of its superior stiffness, strength and thermal properties [16][17]. Additionally, as illustrated by M. Bavdaz et al. [18], Silicon Pore Optics (SPO) is the new X-ray optics technology under development in Europe, forming the ESA baseline technology for the International X-ray Observatory candidate mission studied jointly by ESA, NASA, and JAXA.

As mirror substrates not always provide the desired optical performance, the use of optical coatings to step it up is often required. Coatings has a major impact on the instrument optical performance. Even if mirrors are insensitive to chromatic aberrations, the need of large bandwidth impacts the coating design and the technologies to reach broadband reflective coatings with high reflectivity and low polarization sensitivity. In particular, while most mirrors used for space systems that operate from the ultraviolet (UV) to the infrared (IR) wavelength regions rely on coatings of Ag, Al, Au or Be, extreme regions such as X-Ray, extreme UV and far-IR require specific engineered designs comprising multilayers of different materials. The coating may include adhesion layers (between substrate and reflective layer), interdiffusion layers (between layers of different species) and protection or enhancement layers (on top of the reflective layer or multilayer). Dielectric mirror coatings can be used alone or in combination with metallic ones to prepare multilayer designs. Multilayers of metal-dielectric and all-dielectric films have been extensively used to prepare narrow band reflectors for several spectral bands [19][20]. Multilayers consist in several layers of two or more materials with the right thicknesses to obtain the desired spectral, angular, and/or polarization profile. In the visible and close ranges, multilayers alternate layers of transparent (dielectric) materials, which enable the theoretical design of virtually any arbitrary profile. In ranges such as the extreme UV and the soft x-rays, where the absorption of materials in nature strongly differs from the visible, multilayers may typically alternate a dielectric material and a metal or even two metals.

Both mirror substrates and coatings materials, but also detector technologies and the telescope systems, are all involved in the success or failure of a space observatory. The extreme environment where they must operate implies severe issues in terms of stability and resistance.

**2. Degradation of materials in space - Stability issues on mirrors**

Common to all orbits is the degradation of materials by the hazardous space environment, whose importance in space technology is undeniable [21][22]. Degradation may be caused by atomic oxygen, thermal stress, electromagnetic radiation, telescope outgassing or self-contamination, charged particles, space debris and micro-meteorites. In Low Earth Orbits (LEOs), atomic oxygen (AO) is the main source of degradation, while in the interplanetary medium, the solar wind and solar electromagnetic radiation dominate the degradation effects. Most of the materials used for space optics need to be evaluated for their behavior under several of the aforementioned degradation mechanisms. It is known that these degradation mechanisms can significantly degrade materials and lead to changes in their mechanical behavior or thermo-optical properties [21]. These changes can cause early failures of satellite components or even failures of complete space missions.

The main challenge in the assessment of degradation of materials in space is in the development and choice of the most representative ground testing and extrapolation to end of-life conditions for thermal cycle and for charged particles, AO, UV irradiation, and high-velocity impacts of microparticles. These tests have to account for the different environments in which the mirrors will operate, ranging from Low-Earth Orbit (LEO), to interplanetary orbits and deep space.

Investigations on the behavior of optical materials and coatings in space environment had been reported in a large number of papers starting in the 1970's. Pre-launch acceptance testing and evaluation of mirrors coated for use in space are almost never performed on the actual flight mirror. Smaller witness mirrors, coated at the same time as the flight component, are used as test proxies for the spaceflight component. The intent of the acceptance testing generally aims at identifying any mirror surface quality problem before performing the qualification testing of the final and larger mirror, when recovery from a flaw can be costly. The use of tests samples to verify the performances of the whole mirror is even more important for complex optical coatings such as reflective multilayers [23,24]. Environmental tests are performed to check the resistance of a mirror coating that is exposed to ambient conditions simulating the space environment for the instrument lifetime. As an example, Fig. 1 shows reflectance degradation as a function of wavelength in the UV-Vis spectral range of protected Ag mirrors under various degradation sources. This combination of environmental resistance tests helps to predict, model, and account for the in-orbit degradation of the optical system.

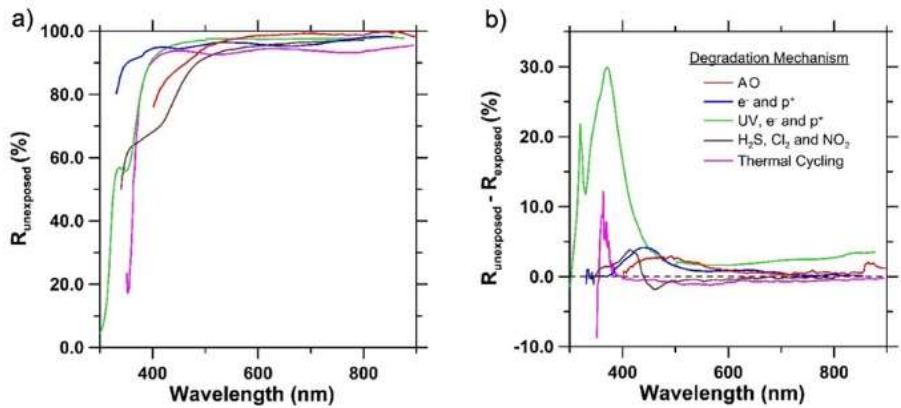

Figure 1. a) Reflectivity of protected Ag mirrors from various sources, unexposed. The differences in the mirror spectral performance is explained by the differences in composition and thickness of the protective coatings on the Ag layer. b) Effect of several degradation mechanisms on the reflectivity of protected Ag mirrors. Red curve: Degradation after 37h of exposure to 5eV AO beam ($4\times10^{20}$ atom/cm$^2$) [25]. Blue curve: Degradation after 279h of simultaneous exposure to 10 keV electrons ($5.3\times10^{15}$ e$^-$/cm$^2$), 2 keV protons ($3.5\times10^{14}$ p$^+$/cm$^2$), and 5 keV protons ($3.4\times10^{14}$ p$^+$/cm$^2$) [26]. Green curve: Degradation after 1436h of simultaneous exposure to solar-equivalent UV, 10 keV electrons ($1.4\times10^{18}$ e$^-$/cm$^2$), and 5 keV protons ($1.6\times10^{17}$ p$^+$/cm$^2$). These dosage levels are equivalent to the radiation exposure at the L2 orbit location over 5 years mission lifetime [27]. Brown curve: Degradation after 240h of simultaneous exposure to purified air mixed with Cl$_2$ (10 ppb), H$_2$S (10 ppb), and NO$_2$ (200 ppb), at 30°C and 70% relative humidity. These conditions are fairly similar to pre-launch environments [28]. Magenta curve: Degradation after 30 thermal cycles from -80°C to +35°C [29].

For each of the key degradation sources (i.e. AO, UV radiation, thermal cycling, charged particles, telescope outgassing, space debris and dust several mitigation techniques and strategies have been proposed, most of them based on coatings. Coatings are in general a fundamental technology for space science and applications, and degradation and radiation issues have been extensively investigated since the beginning of space programs. In particular, coatings performing critical optical functions have been used in space instrument applications for NASA, ESA and the other international and national space agencies for more than 50 years. The performance of the first coatings launched into space had been observed to change with time. Starting from that, pre-flight testing in simulated space environments have been developed to verify spectral and efficiency performance, which desirably predict the changes observed in space. The effect of real or simulated space conditions on mirrors has been investigated during the last decades and in the following sections we will discuss the main results and developments reported in literature.

The next subsections address the main degradation sources in space environment. A large emphasis is given to the far UV (FUV, $\lambda$ in the 100-200 nm) and the extreme UV (EUV, $\lambda$ in the 50-100 nm), due to the enhanced degradation that arises due to the strong absorption of most contaminants in these ranges compared to longer wavelengths.

*2.1 Atomic Oxygen*

AO is the main atmospheric component in LEO up to altitudes of 700 km. It is a species with large harmful potential over many materials. The intrinsic reactive capacity of AO as a free radical of a very electronegative element adds to the strong relative velocity between the orbiting spacecraft and the thermal distribution of orbital AO, which strengthens oxygen capacity to react with and to sputter off the target material. It is also an indirect source of contamination, as its interaction with organic materials, such as polymers, may originate secondary volatile compounds which in turn might condensate on critical elements of the telescope, such as on optical surfaces. Optical surfaces are degraded in a level directly proportional to AO fluence. This, in turn, is determined by several factors, including [30]: spacecraft altitude, as AO decreases with altitude; optical surfaces orientation, as surfaces in the ram or windward direction will be exposed the most; orbital inclination, as high inclination orbits expose optics to cosmic radiation which in turn may increase the AO generation and hence exposure; solar activity, as the Sun emits radiation and charged particles that can promote the generation of AO; and mission duration. The degradation issues caused by the impact of AO in the space environment has been investigated by several authors [31]. It is particularly harmful in LEO, where AO is formed by molecular oxygen dissociation through solar UV radiation at altitudes greater than 100 km. When combined with typical spacecraft orbital velocities of several km/sec, it has the effect of exposing the optical system to a stream of AO at an energy of approximately 5 eV. Hence, optical components intended to operate in LEOs need to be designed to resist atomic oxygen. Nowadays most of the flight optics undergo a critical 5-eV energy AO test for their space qualification, where the AO total fluence and exposure time on the coatings is typically calculated from numerical models and intended to mimic the extent of the entire mission [32].

While most of the oxide-based substrates are resistant to AO, bare metal surfaces and coatings may be vulnerable. The EOIM-III experiment tested the resistance of several optical materials to AO during the Space Shuttle mission 46 [33]. Among the most interesting results, coating materials such as fluorides ($MgF_2$, $CaF_2$ and LiF) and Ir and Pt showed no significant damage, but Ni mirrors showed oxide formation and the reflectivity of Au mirrors overcoated with Ni was decreased by the degradation of Ni. W. Duan et al. [34] investigated the effect of space AO on the polarization contrast of polarization modulated mirrors under different experimental doses by using a terrestrial simulator. Peters et al. [35] exposed Os, C, and bare Ag to ambient AO in a space shuttle flight. Post-flight laboratory analysis revealed that the unshielded C and Os films were totally removed, presumably by formation of volatile oxides. Bare Ag was drastically modified to a nonconductor. A later attempt to protect Os with a 6-nm thick Pt film did not resist exposure to AO in the orbital direction and volatile Os oxide escaped through gaps in the Pt film, which resulted in a poor UV reflectance measurements on both the unexposed as well as the exposed areas; on the other hand, a 10-nm thick film of Pt provided almost complete protection [36]. However, such protection thickness would hinder the relatively large EUV reflectance of Os. Peters et al. [37] exposed films of various metals to a long LEO exposure. All materials, Cu, Ni, Pt, Au, Sn, Mo and W, were somewhat affected by oxidation with AO, mostly in the ram direction of the spacecraft, although they were not affected as severely as it had been found for Os, C and Ag. Oxidation ranged between Au, the most stable, and Cu, the most affected. Another experience to protect Os in order to avoid AO attack was carried out by Hemphill et al. [38]. A 2-to-3-nm thick Ir film was seen to protect an Os film, which had been deposited either on a Rh film or on a second Ir film. Such 3-layer structure preserved the Os high EUV reflectance characteristic at grazing incidence to be used on gratings in the 9-26-nm spectral range. The effect of LEO AO on C was also analyzed by Hadaway et al. [39], who exposed diamond-like C (along with 12 other materials) to LEO environment and measured the total integrated scattering in situ over time. After several weeks, the C film was completely eroded away. Gull et al. [40] exposed films of Os, Cr, Pt, and Ir to the LEO environment for a few days and its effect on EUV reflectance was measured. Os was the most severely affected when exposed to the ram direction, where it disappeared, whereas there was little change when it was masked. Cr, Pt, and Ir were much less affected. Ir underwent some reflectance decrease at wavelengths longer than 160 nm. Pt increased reflectance after exposure, which was attributed to the cleaning effect of AO on a sample that was assumed to be previously contaminated. As mentioned above, the presence of AO on the orbit may not only degrade the coating, but it also has the potential to remove contaminants from various sorts of coatings.

Herzig et al. [41] also exposed transition-metal mirrors of Au, Ir, Os, and Pt to LEO environment, close to the ram direction. As with the aforementioned experiments, Os disappeared, whereas Pt and Ir behaved relatively well after exposure. Au suffered a severe reflectance decrease, but, even though some outer monolayers may have been sputtered off, the decrease was attributed to contamination from the surrounding areas. The same authors also exposed chemical-vapor-deposited (CVD) SiC to LEO environment and found that its EUV-FUV reflectance was severely affected, and the degradation was much larger for the exposed area than for a masked area. Degradation was attributed to surface oxidation to $SiO_2$. The effect of AO on CVD-SiC EUV-FUV reflectance and the synergic effect of AO along with UV radiation on the CVD-SiC near-UV reflectance were reported by G. Raikar et al. [42] and S. Mileti et al. [43], respectively. The loss in performance does not exclude the use of CVD-SiC for missions where oxygen in not present. Other than high temperature CVD-SiC, carbides deposited by sputtering at room temperature are a choice of moderate EUV-reflectance mirror that is attractive for optical coatings [44,45]. Keski-Kuha et al. [46] tested the ability of ion-beam-sputtered deposited SiC and $B_4C$ to withstand the exposure to the LEO AO. For SiC, a severe reflectance decrease was observed when the coating was oriented in the ram direction and 3 more times silicon oxide was found compared with the control sample that was kept in the lab, which was attributed to the direct exposure to AO. A second SiC sample was exposed to LEO AO but it was placed at 160° from the ram direction, so that it was protected from the effects of direct AO bombardment. This sample displayed only a slight reflectance degradation, typical of an aged sample. Three $B_4C$ samples were also exposed to LEO AO at 0°, 26°, and 160° from the ram direction. All three samples experienced some EUV reflectance reduction, larger than the typical sample ageing, but reduction was not as drastic as with SiC. The extra $B_4C$ reflectance reduction was mostly attributed to contamination. No roughness increase was observed for either SiC or $B_4C$.

Herzig et al. [41] flew $Al/MgF_2$ mirrors and exposed them to LEO environment. Even though some samples kept their FUV reflectance, one sample experienced significant reflectance degradation in the range around 150 nm, and smaller degradation was produced at a wavelength of ~120 nm or 200 nm. Degradation could not be attributed to contamination, since the largest sensitivity to contamination was expected at ~200 nm, where 25-nm thick $MgF_2$ layer happens to be a quarterwave thick. The change at 150 nm was then attributed to plasma resonance absorption in Al induced by surface roughness, even though no significant difference in roughness before and after orbit exposure was observed.

To reduce or eliminate atomic oxygen erosion of materials in space, the application of thin-film protective coatings made of durable materials is the most used approach [25,26,47–51]. As previously described, oxides and fluorides are materials relatively resistant to AO, making them suitable as a capping layers in coatings for space optics. For example, I. Gouzman et al. reported on the durability of protected silver surfaces in an AO environment [25]. In this case the protective layer consisted of a thin $Al_2O_3$ film, as alumina has been considered one of the suitable material choice to be applied as protective coating. Interestingly they applied two approaches to test AO resistance: radio-frequency (RF) oxygen plasma exposure and laser detonation source of 5 eV AO. It was suggested that the RF plasma environment is too severe for realistic simulation of the AO interaction while a 5 eV AO exposure demonstrated that the protective coating was suitable for potential LEO applications. Silicon Dioxide ($SiO_2$) and Magnesium fluoride ($MgF_2$) are other commonly used coatings in the vacuum ultraviolet (VUV) spectral region because of their high reflectivity down to 110 nm. $MgF_2$ coating, for example, is used on Hubble Space Telescope optics covering the wavelength range from 110 nm to near infrared. Even though quite effective, $MgF_2$ protected aluminum is a soft coating that scratches easily [3]. Therefore optical components with this coating have to be handled carefully to avoid damage to the coating. Lithium fluoride can extend the useful range of aluminum down to the LiF absorption cutoff of 102.5 nm. Unlike $MgF_2$ characterized by excellent protection capability and long life, LiF thin films are hygroscopic and exhibit reflectance degradation and increased scatter with age. Al high intrinsic reflectance extends beyond $MgF_2$ and LiF cutoff wavelengths down to ~83 nm. However, Al reactivity in presence of oxygen results in a dramatic FUV/EUV reflectance decrease and no transparent material is available in nature to preserve reflectance to such a short wavelength. The degradation of FUV reflectance of unprotected Al through controlled oxidation to $O_2$, $H_2O$, and other species[52][53] and to AO [54] has been investigated. AO was found to be orders of magnitude more effective to degrade Al reflectance compared with the same doses of $O_2$. Non-protected Al mirrors have been also exposed to LEO environment [41]; even though Al oxidation occurs rapidly, which had happened right after the sample was taken out of the vacuum chamber in the lab, Al mirrors experienced further reflectance losses below 250 nm once in orbit, which was attributed to a greater reactivity of Al with AO compared to atmospheric $O_2$. In view of the sensitivity of bare Al to react with AO, some procedure to significantly reduce the rate of impingement of oxidizing species must be developed, either based on going to high enough orbits [55] or through the use of some scheme that shields the mirrors from ambient oxygen [56][57]. Al extended reflectance was not ignored by space instrument developers since early stage in the space era, and such an extension would be very beneficial for space observations due to the presence of important spectral lines in the extended region. Especially one main difficulty was envisaged: how to keep a bare Al coating free of contaminants, mainly free of oxygen. Two strategies were then conceived: i) Depositing the Al coating when the optical instrument is already in orbit and after a long-enough time for a thorough spacecraft outgassing; ii) Depositing the Al coating on Earth together with some volatile material to prevent Al deterioration, a material that could be removed once in orbit through some mild bakeout. The main impulse for the latter strategy was named REVAP [58]. Among the two mentioned strategies, the in-space aluminization has been addressed in a higher number of proposals, with the first studies dating back to the mid 1960's. A demonstration to aluminize mirrors in space was performed in 1975 onboard of the Orbiting Solar Telescope in the Soviet space station Salyut 4 on a 337-350-km altitude orbit. Even though the process of Al evaporation worked well, no UV reflectance increase was observed after two aluminization processes, probably due to fast oxidation by ambient AO [59]. Several other experiments have been considered or proposed [58][60] but, to the best of our knowledge, no other experiment to aluminize mirrors in space have been carried out.

*2.2 Thermal processes*

Thermal cycling may cause mechanical defects that can grow and degrade the optical system performance on orbit. Thermal fluctuations may, for instance, induce mechanical stress that may lead to alterations in the figure of optics [61], or modify the stress balance between the coating and substrate, or even between different materials within the coating. Nowadays most of the flight optics undergo a critical thermal cycling test for their space qualification. This test exposes optics to a one or more cycles over temperature ranges typically within [-100°C, +100°C] for 24 h or more, although for some missions this test might be more extreme. As a reference, MIL-M-13508C specifies that protection Al coatings located in front mirrors have to survive at least 5h at -62 °C and 5h at 71 °C.

One example of an extreme temperature range test was the coating qualification of the oxide-protected Au-coated Be mirrors for JWST, in which witness samples were cryogenically cycled to down to 15 K four times and to 328 K one time [62].

Among others, R. K. Banyal et al. reported on thermal characteristics of a classical solar telescope primary mirror [63] (similar investigations have been reported by L. Rong et al. [64]). They used a heat transfer model that considers the heating caused by a smooth and gradual increase of the solar flux during the day-time observations and cooling resulting from the exponentially decaying ambient temperature at night. The thermal and structural response of SiC and Zerodur was investigated in detail. The low thermal conductivity of Zerodur mirror gives rise to strong radial and axial temperature gradients for the day-time heating and night-time cooling. Heat loss by free convection is very slow so the mirror retains significant heat during the night. The observed thermal response of the SiC mirror is significantly different from Zerodur. The temperature within the SiC mirror substrate equilibrates rather quickly due to high thermal conductivity. The thermal expansion of ceramic, silicon and SiC optical substrate materials was also investigated in regard to Herschel (2009-2013) observatory [15]. In particular, SiC is one of the most investigated materials for an observatory in cryogenic environment [65][66][8,63][67].

Research on coatings and thin films demonstrated that the instability of properties in optical film was attributed to the coating materials and their deposition process [26,29][50,68][69][70]. For example, with respect to metals, metal oxide compound coating materials possess large energy gaps and provide high transmission to short, near-UV wavelengths because their optical absorption edge is outside (shorter than) the wavelength of interest. Therefore, they are intrinsically less vulnerable to damaging by thermal effect, ionizing and UV radiation. The most commonly used coating materials are $MgF_2$, $ZrO_2$, $TaO_5$, $TiO_2$, $HfO_2$, and $SiO_2$ [71].

*2.3 Ultraviolet Radiation*

UV radiation comprises the spectral range of wavelengths between few nanometers up to 400 nm. The effects of high energy photons on mirrors are not strictly related to the reflectivity or morphological properties. The effects from these photons are not the determining factor contributing to radiation damage. However, chemical changes such as reduction and oxidation reactions can induce optical absorption in thin film layers, and UV photons can promote such reactions, changing the composition of the materials. For these reasons, space UV and ionizing radiation durability of materials must be considered. Importantly, the radiation effects are synergistic with other effects and must be considered together [72]. One of the principal effects of UV radiation is the polymerization and darkening of silicones and hydrocarbons, which are ubiquitous contaminants in space telescopes. This darkening effect is often enhanced by electron irradiation [73]. Hence, the UV resistance of mirrors is often tested during space qualification tests. It is common to use a distribution of Xe lamps (or similar sources) to obtain a spectral intensity profile similar to the solar irradiance, and the mirrors are exposed for a time equivalent to the intended operation hours under solar ultraviolet exposure [74].

The first space optical coatings used for band-pass filters were constructed of thermally-evaporated soft materials such as ZnS and $MgF_2$. Exposure to the space environment containing ionizing radiation, solar UV, atomic oxygen and high vacuum revealed the unstable operation of those coatings. In addition to humid-vacuum shifts in wavelength properties, filters, anti-reflective (AR) coatings and other coatings suffered radiation-induced transmission loss that was especially pronounced at short wavelengths. UV exposure may have effects on polymers and other materials used in lightweight mirror material in spacecraft applications. In this latter case, the effects of UV exposure need to be accounted for due to their potential impacts on the thermal management of a spacecraft and during application in composite mirror structures [75][76,77].

ZnS deposited by evaporation was a coating material used decades ago for its moderate FUV reflectance and its transparency above 400 nm. Hass et al. [78] evaluated the resistance of a ZnS film to intense UV irradiation as it would be expected in a space instrument. ZnS experienced a dramatic reflectance decrease in the UV after a long UV irradiation in air, whereas reflectance decrease was relatively small longwards of 400 nm. The outermost 15-20 nm thickness was seen to have changed from polycrystalline ZnS to amorphous ZnO. The authors also studied a multilayer with single Al and Ge films under outermost ZnS film to enhance FUV reflectance and to decrease near UV and visible reflectance. The Ge/Al/ZnS multilayer was UV irradiated in vacuum, which resulted in a severe FUV reflectance decrease, increasingly more severe towards shorter wavelengths, whereas no change was observed longwards of 260 nm. Again, there was a predominant presence of oxygen over sulphur in the outermost 10-20 nm. The paper reported that, even in the total absence of oxygen upon UV irradiation, sulphur is expected to sublime, leaving a metallic film of Zn. All this behavior recommends caution in employing ZnS as the outer coating of an optics in space.

Fuqua at al. [79] reported the on-orbit degradation of Ag mirrors on the Suomi-NPP spacecraft. They sensed an important degradation in near-IR bands of the Visible Infrared Imaging Radiometer Suite instrument, but little degradation in the green and blue channels. They first considered the possibility that the mirrors had become contaminated either before launch or on-orbit, and that the contaminant was darkening with UV exposure. However, the spectral signature of the degradation was uncharacteristic of UV darkened molecular contamination, which typically results in greater losses in the short wavelengths rather than the NIR. After an investigation on flight witness mirrors, they concluded that a non-qualified process was employed in the production of the flight mirrors, which inadvertently deposited a thin layer of tungsten oxide, WOx, on the surface of the mirrors. The tungsten oxide, when illuminated with UV, becomes absorptive in the near infrared with a spectral dependence that compared very well with the inferred behavior of the mirrors on orbit.

*2.4 Outgassing and cross-contamination*

As previously mentioned, one of the main contamination sources in space mirrors originates from outgassing in the space vacuum environment. Due to the strong absorption of materials, particularly contaminants in the FUV range, space contamination mainly originated in the vehicle and instrument outgassing has been investigated by several authors and the FUV properties of most volatile spacecraft materials have been measured [55][80–84], which indicate larger absorption in the FUV compared to longer wavelengths. These experiments are useful to evaluate the top allowable contaminant thickness before FUV reflectance is unacceptably degraded. When the coating is irradiated with strong UV radiation, this results in that the contaminant is transformed through a photopolymerization process and becomes non-volatile, so that it gets stuck to the coating. UV radiation provides the energy to break bonds in the hydrocarbon chain and stimulates intermolecular crosslinking [55]. The photopolymerization process mostly depends on the coating and contaminant nature, on substrate temperature and on the specific UV radiation energy and intensity [85]. A facility was developed at GSFC to controllably contaminate mirrors and in situ measure their FUV degradation [85]. It is important to point out that UV irradiation with no contamination does not degrade Al, Ag or Au based mirrors (e.g. Al/MgF$_2$ reflectance [86]), but the combination of both does. Other than UV, energetic protons and electrons may also contribute to turn a contaminant into nonvolatile [55].

A strong manifestation of the concomitance between contamination in space and UV was observed after the first servicing mission on HST [87]. The Wide Field and Planetary Camera I (WFPC-1) was replaced and returned to Earth. The WFPC-1 pickoff mirror was analyzed. The Al/MgF$_2$ mirror was found to be covered with a 45-nm thick contaminant, which severely degraded FUV reflectance. The contamination was attributed to the outgassing of HST during its first 3.5 years operation. The mirror was found to be contaminated with hydrocarbons, esters, and silicones. Figure 2 shows drastic reduction in reflectivity, and x-ray photoemission spectroscopy data on the surface of the returned mirror revealing the composition of the contaminants. The mirror was carefully cleaned and the preflight reflectance was fully restored, so that it was demonstrated that the Al/MgF$_2$ coating had not been degraded [87]

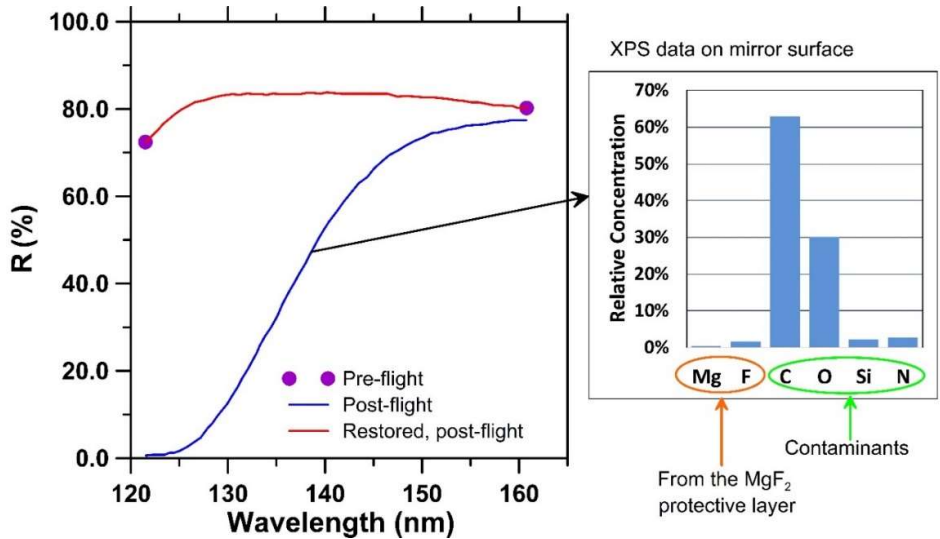

Figure 2. Evolution of the far-ultraviolet reflectance of the Wide Field and Planetary Camera-1 (WFPC-1) pick-off mirror (based on Al protected with MgF$_2$). Purple points: pre-flight data. Blue: post-flight data after 3.5 years of deployment in space, with a severe reflectance degradation. Red: reflectance recovery after contamination removal with a chemical cleaning. The inset on the right depicts XPS data acquired on the surface of the recovered mirror after its return to Earth, showing the presence of contaminants such as C, O, Si, and N. [88]

A later servicing mission enabled retrieve other Al/MgF$_2$ mirrors from HST instruments for inspection after 15.5 years in space [89]. While two COSTAR optics mirrors kept a relatively high FUV reflectance, comparable to or even better than a witness sample that had been stored in a desiccator, WFPC-2 pick off mirror resulted in a reflectance degradation as severe as for the aforementioned WFPC-1 mirror. This suggested a similar contamination for both mirrors, in spite of the efforts carried out to reduce contamination on WFPC-2 after the experience with WFPC-1. The different levels of contamination through the mirrors were unexpected and attributed to contamination dependent on the specific location within HST hub.

Regarding grazing-incidence mirrors, Osantowski calculated the sensitivity of mirror reflectance to a range of optical constants selected for generic contaminants, such as hydrocarbons [90]. Three wavelengths were investigated, 10, 50, and 100 nm, as representative of the EUV. He calculated critical contaminant thicknesses to reach allowable reflectance changes. A preliminary conclusion was that Au and Zerodur mirrors are relatively insensitive to surface films, which can even result in an increased reflectance in some cases. Mrowka et al. investigated the effect of intentional contamination of grazing incidence Au mirrors with vacuum pump oil to evaluate the allowable reflectance decrease by contaminants of an instrument to be part of EUVE space telescope [91]. To check the effect of contamination with a common contaminant, a coating with 15-nm oil resulted in a total recovery of the original EUV reflectance after a long-enough outgassing time in the reflectometer vacuum chamber, with no increase of scattering either. But when a 50-nm thick oil was deposited, outgassing reduced such thickness just to 35 nm, and mirrors kept a hazy look. Since the remaining oil deposit was known to be in droplet form, an increased scattering for the coating was expected. Polymerization was discarded because the estimates of UV irradiation and charged particle fluxes were too small.

*2.5 Charged particles*

An additional concern in space optics regards the degradation occurring on mirrors when they are exposed to charged particles and ions. During an inter-planetary journey, galactic cosmic rays background and Sun are the main sources of such particles and ions. Galactic Cosmic Rays (GCRs) are a continuous and isotropic flow of charged particles reaching the solar system from outside the heliosphere. They are approximately composed by 85% of protons, 14% of helium, and the residual 1% of heavy ions. The energy spectrum ranges from few MeV up to GeV with particles fluxes that decrease with increasing energy. Inside the heliosphere, GCRs decreases by a few %/Astronomic Unit (AU) with heliocentric distance (R) and the solar activity modifies the GCRs flux. As the solar activity undergoes the 11-year cycle, the GCRs flux varies with the maximum during solar minimum periods [92]. Sun emits particles such as protons, electrons, alpha particles and less abundant heavy ions such as $O^{+6}$ and $Fe^{+10}$ continuously (solar wind) or as part of eruptions (unpredictable occurrences) or coronal mass ejections. The solar wind is an outflow of completely ionized gas originating from the solar corona and expanding outwards to the interplanetary regions. Different components are contained in the solar wind, which differ for particles speed, spectral flux (particles / eV cm² s) of the constituents and solar region of provenience. For instance, the "quiet" solar wind in the ecliptic plane is constituted by protons of ~1 keV and alpha particles of ~4 keV whereas out of the ecliptic such energies can increase up to 4 times [93]. More severe but transient disturbances can be caused by energetic particles events occurring during coronal mass ejections or solar flares. Such events can potentially lead to high fluxes of protons in the energy range from tens to hundreds of MeV, whose effects can be occasionally detected even at Earth ground. This proton emission occurs randomly and usually during periods of solar maxima, and it is accompanied by heavy ions. In general, the fluence of solar energetic particles scales with distance from the Sun as $R^{-3}$ at a few MeV and $R^{-2}$ at tens of MeV and above [94].

Around planets, the space environment is also affected by their magnetosphere, which interacts with charged particles present in the heliosphere. Moreover, albedo neutrons generated by GCRs interaction with the planet atmosphere decay into protons, giving an additional source of ions around planets. These particles are confined via magnetic mirroring and trapped preferably in some regions around the planets, forming radiation belts [95]. For example, Earth has two main electrons belts at about 3000 km and 25000 km of altitude with energies varying from few keV up to 10 MeV; protons are instead confined in a belt at around 3000 km of altitude in which the energies span between 100 keV and several hundred MeV. Outside these radiation belts the distribution and flux of particles depend on the characteristics of the magnetosphere, the planetary atmosphere, the Sun distance and the phase of the solar cycle. Earth geostationary orbits (GEO; circular orbits at 35786 km altitude) has an electron flux ranging between $10^9$ e/(cm²s) and $10^8$ e/(cm²s) in the energy interval 1-10 keV and $10^5$ e/(cm²s) at 1 MeV. The proton fluence in the same orbit is $10^{10}$ p/(cm²s) at 1 MeV and decreases by two orders of magnitude at 10 MeV and four at 100 MeV. The magnetosphere of giant planets, such as Jupiter, becomes an important source of high-energy electrons (>10 MeV) in the interplanetary space [96].

The spacecraft components need to be protected by highly penetrating radiation and particles encountered in the operational environment. In fact, highly energetic photons as well as MeV particles can easily penetrate mm thicknesses of materials, undergoing a deceleration in case of particles and, in general, producing secondary photon and particle emissions. By its nature, secondary particles have to be analyzed on a case-by-case basis through Monte-Carlo simulation in order to obtain global information that can be used during the design and testing procedures. For this reason, spacecraft requirements always include a total ionizing dose (TID) specification (expressed in krad), a value that corresponds to the total energy deposited in matter by ionizing radiation per unit of mass. By definition, TID is an integral dose, and therefore it takes into account the cumulative effect due to particles of different energies. The ground validation of the spacecraft components is then usually performed by evaluating the effect given by a specific TID, deposited via acceleration facilities. Although this approach is suitable for testing the radiation-hardness of an electronic component or investigating the degradation of the opto-mechanical properties of bulk materials, it becomes inappropriate for the optical coatings because the effects occurring in the thin films strongly depend on the proton energy and therefore the implantation depth of particles. High-energy particles penetrate deeper in the optical components, in the order of tens of μm or more, interacting very little with the nanometric coatings and depositing all the energy in the substrate. In contrast, keV ions implants within the coatings with a profile highly dependent on their density, potentially inducing changes of their optical, structural and morphological properties. As a general rule, we can affirm that thin films in the nanometric scale are mostly affected by low energy particles, that implant in the coating itself, but not by MeV particles, which overcome the structure, eventually reaching the substrate. Experiments with MeV electrons and protons with typical fluences faced in the space environment (i.e. < $10^{12}$ ions/cm$^2$) have proven negligible degradation effects on optical coatings having a total thickness lower than few microns in the visible-UV [97][98][99], in FUV [86] and even in the EUV [100]. Visible multilayer filters irradiated with protons at 4, 18 MeV and 30 MeV [97] [99] and electrons at 50 MeV [98] showed no changes after irradiation. Canfield et al. [86] irradiated Al/MgF$_2$ mirrors with 1-Mev electrons and 5-Mev protons. No effect on the reflectance at 121.6 nm was observed. Hass and Hunter [55] reported also the effect of energetic electrons and protons on Al/MgF$_2$ coatings.

The investigation of the effects induced on optical mirrors by low-energy particles and ions are performed by using terrestrial facilities based on ion implanters and accelerators. However, simulation of the space environment exposure is extremely challenging since it is impossible to approach the irradiation conditions occurring in space. For example, while the exposure in space typically lasts for several years, a ground-based experiment needs high particle flux rates in order to reach the mission life-time expected fluences in a reasonable amount of time. Moreover, during the accelerating tests potential synergistic effects, not present in space, can arise, such as thermal effects related to the high flux and surface contaminations due to the contaminants present in the vacuum chamber [101]. Moreover, irradiation experiments are also highly influenced by practical reasons, such as the availability of a facility able to provide a specific ion species, energy and flux.

In the case of low energy particles, damage depends on energy, flux and fluence. Low energy proton irradiations (< 500 keV) with fluences lower than $10^{16}$ p/cm$^2$ have shown negligible changes in the near infrared reflectance of SiO$_2$-protected Al mirrors [102][103]. A degradation dependent on the proton fluence have been instead observed in the visible and near ultraviolet. For example, Hai et al. and Qiang et al., [104][105][103][106] investigated the effect of 60-keV protons on Al protected with SiO$_2$. These mirrors were measured from the near UV to the IR. Reflectance was monotonously degraded with proton dose (see Figure 3a): with a fluence of $10^{16}$ p/cm$^2$, a reflectance drop of 5% at 700 nm and 10% at 500 nm was observed, whereas in the UV this drop goes over 20%. Moreover, fluences in the order of $10^{17}$ p/cm$^2$ heavily compromise the Al-protected mirror reflectance up to over 1000 nm. Similarly, Al/MgF$_2$ mirrors of unknown design but optimized for the near UV or visible, hence with a thicker MgF$_2$ protective coating than FUV mirrors, were exposed to a geostationary orbit simulator consisting in simultaneous irradiation with UV, electrons and protons [72]. UV reflectance decay depended on the specific mirror, which had been prepared by a specific vendor, so that some degradation could not be discarded for the mirrors in space environment. Such degradation may not have been due to the presence of contaminants but to the shallow penetration of electrons and even more to less penetrating protons. Such reflectance decrease was attributed to the change of Al optical constants or to the appearance of ripples and hillocks on the surface of the Al mirror. Calculations on the effect of protons over general metallic surfaces predict the recombination of protons to form H$_2$ bubbles; these bubbles will result in a significant roughness increase. The proton energy used in the irradiation experiment greatly influences the degradation results. For example, based on the results reported in [103] it can be observed that a fluence of $10^{17}$ p/cm$^2$ induces a higher reflectance degradation when the proton energy is low. This fluence at 60keV induce a reflectance drop of 99% at 400 nm while at 160 keV this drop is about 20% (see Figure 3b). This degradation is due to the different ion implantation profile inside the coating: the lower is the energy the shallower is the ion implantation peak. In case of metallic mirrors, if the implantation peak falls in the topmost part of the metallic layer or inside an eventual protective layer, the bubble formation inside the coating will provide a greater degradation.

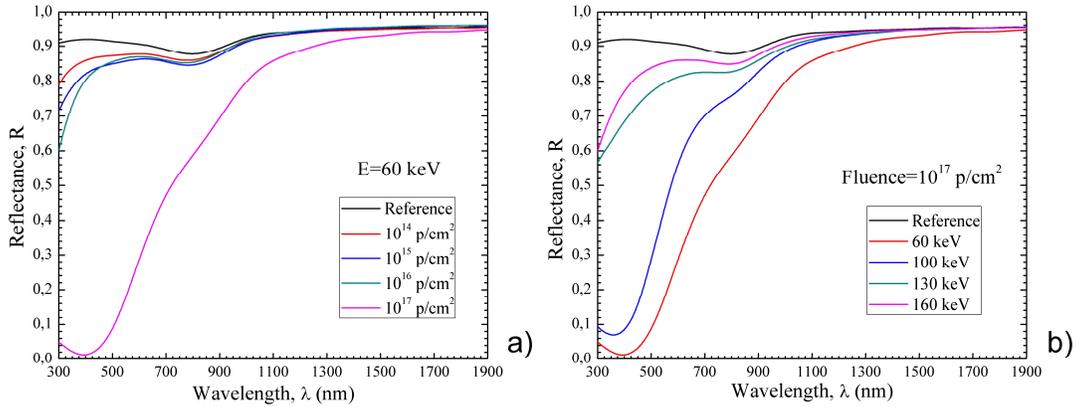

Figure 3: a) Evolution of the UV-VIS reflectance of a SiO$_2$-protected Al mirror irradiated with protons at 60 keV with different fluences (data retrieved from [103]). b) Evolution of the UV-VIS reflectance of SiO$_2$-protected Al mirror irradiated with protons at different energies and keeping a fluence of $1\cdot10^{17}$ p/cm$^2$ (data retrieved from [103]).

Gillette and Kenyon [107] exposed Al/MgF$_2$ and Al/LiF FUV mirrors to 10-keV protons to simulate several-year exposure in a synchronous earth orbit. Such irradiation resulted in a broadband reflectance decrease centered at ~210 nm (Al/MgF$_2$) and ~190 nm (Al/LiF). Reflectance decrease grew with proton dose. Reflectance decrease was negligible at the short-end of the high FUV reflectance range of these mirrors, which was due to the presence of contamination but not coating degradation. The contamination thickness was calculated to be 4-5 nm. Even though an undulating pattern on the coating surface was induced by irradiation, its small width did not result in observable scattering. Most reflectance degradation could be reverted to close to the original reflectance after exposing both Al/MgF$_2$ and Al/LiF mirrors to AO, which was attributed to oxidation of the contaminant to turn it volatile. Gillette and Kenyon [107] exposed Pt mirrors to 10-keV protons to simulate a long exposure in a synchronous earth orbit. Reflectance degradation in the full 93-250 nm measured range presented no spectral structure, which may be due to the lack of interferences, contrary to what was observed for Al/MgF$_2$ and Al/LiF mirrors.

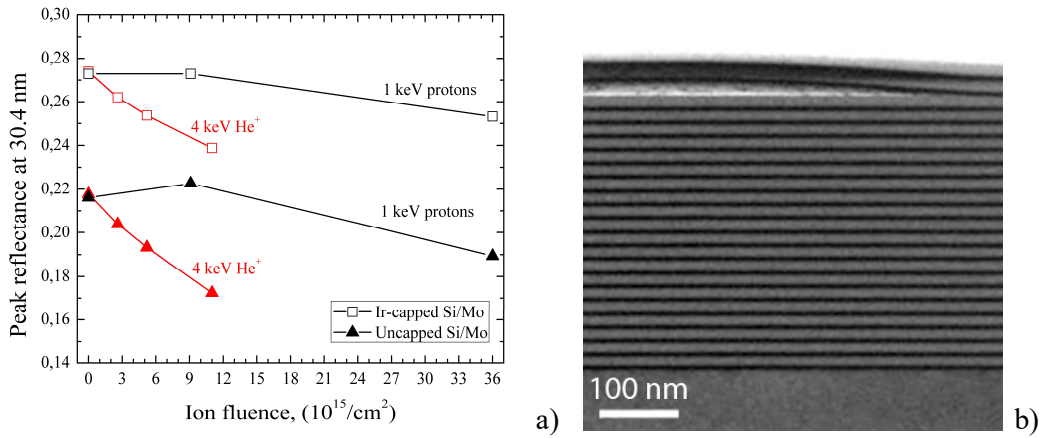

Figure 4: a) Evolution of the EUV reflectance at 30.4 nm of an un-capped and Ir-capped Si/Mo multilayer coating versus 1 keV protons ad 4 keV He ions fluence (data from ref. [108][109]). b) Delamination occurring on a Si/Mo multilayer under a 1 keV proton irradiation with a fluence of $3.6\cdot10^{16}$ p/cm$^2$ (image from Ref. [108]).

A similar behavior was observed with He ions. Low energy He ion irradiations on metallic thin films of Au and Cu demonstrate that with fluences of the order of $10^{15}$ - $10^{16}$ ion/cm$^2$, a faint dislocation band starts forming, with preservation of the optical performance in the visible spectral range and a fluence-proportional degradation in the ultraviolet [110]. Fluences of the order of $10^{17}$ ions/cm$^2$ induce a large formation of bubbles inside the films and a deep transformation of the surface morphology [111][112] with a consequent degradation of the visible and UV reflectance. The diameter and the density of such bubbles increase with the fluence due to the tendency of helium ions to migrate form agglomerates. This behavior has been confirmed not only in metals, but also in semiconductors [113].

A particular case is instead that of ML stacks for the EUV. Several studies have demonstrated that protons and alpha particles with energy of few keV can already lead to dramatic degradation of performance with fluences in the order of $10^{16}$ ions/cm$^2$ [108][109][114]. For example, Mo/Si structures with different capping-layers were irradiated by protons at 1 keV with fluences of $9 \cdot 10^{15}$ p/cm$^2$ and $36 \cdot 10^{15}$ p/cm$^2$ showing a change of the peak position and a degradation of the reflectance (Figure 4a). Such effect was attributed to the blistering, expansion and delamination and occurring in the topmost layers of the ML stack (see the TEM image reported in Figure 4b) [115][116][117]. After He$^+$ ion irradiation with fluences of $2.5 \cdot 10^{15}$ ions/cm$^2$, $5 \cdot 10^{15}$ ions/cm$^2$ $10^{16}$ ions/cm$^2$ the exposed MLs show a drop of reflectance but no appreciable reflectance peak shifts [109]. In this case, the degradation was attributed to an increase of the intermixing at the interfaces in the topmost layers.

Recently, Al/Mo/B$_4$C and Al/Mo/SiC were also irradiated with 1 keV and 100 keV protons with doses of $7.4 \cdot 10^{12}$ p/cm$^2$ and $9 \cdot 10^{15}$ p/cm$^2$. The lowest dose was chosen in order to simulate the situation expected inside the High Resolution Imager (HRI) and Full Sun Imager (FSI) telescopes of Solar Orbiter mission, where the mechanical structure and the front filters drastically reduce the proton flux impinging on the multilayers. None of the irradiated structures showed appreciable changes of their performance, suggesting that at these fluences MLs can be considered stable.

*2.6 Dust and Space Debris*

Dust (or meteoroids) and space debris are important sources of mirror degradation. In extreme conditions, meteoroids may cause a full spacecraft failure; one example of the latter was the Olympus communications spacecraft (ESA), in which the general failure of the satellite was attributed with a high probability to a Perseid meteoroid impact[118]. Although dust and debris are small in weight and size (e.g. millimeter- and micron-size particles are the most abundant in LEO), their high velocities, ranging from few m/s up to dozens of km/s, represent a threat for space optics. In the near-Earth, space debris is generated by launch activity and subsequent operational practices, and its flux is comparable with the one of meteoroids in the size range between 10 μm to 1 mm[119]. However, meteoroids are generally more harmful than space debris, as the average velocity of the former is higher. Beyond LEO, space dust is dominant, where short-period comets with aphelia less than 7 AU have been identified as a major source of interplanetary dust released through the sublimation of cometary ices[120]. Aside from the mass and velocity, the effect of these particles can be further exacerbated by the directionality of the optical surfaces relative to the ram direction. The exposure time is also critical; it has been reported that even after a short exposure to the space environment, exposed surfaces can be covered with impacts from small-size debris and meteoroids[121]. The chemical composition of dust is diverse, but mostly are silicates, ice, and iron and organic compounds. Depending on several factors, dust can accumulate on the surface of the optics and increase the scattered radiation[122], or flake the protective coating leaving reactive materials exposed (which could be subsequently degraded by AO, for instance), or it can even blast away the coating and produce craters on the substrate. In extreme conditions, high-velocity collisions may produce plasma, generating side-effects that may be more damaging than the purely mechanical effects[118]. Additionally, the collateral effects of impacts may damage or contaminate optics; an example of the latter will be explained below in the description of one of the NASA Long Duration Exposure Facility (LDEF) experiments.

The experimental verification is often necessary to better understand the effect of high-velocity particles in optics. This can be performed in dedicated testing facilities, such as Heidelberg Dust Accelerator, in which particles of various materials can be accelerated to velocities up to 40km/s.[123] Heaney el al. [124] utilized the aforementioned facility to simulate the effect of the impacts of iron (1.2 µm diameter) and latex (0.75 µm diameter) to mimic inorganic- and organic-based meteoroids, at velocities of 2-20 km/s on an oxide-protected, Au-coated Be witness mirrors for the JWST. He found that both latex and iron particles had the ability to blast away the protective coating, creating craters in which the Be substrate was exposed. He reported ratios between the crater diameter and the incident particle kinetic energy of 0.09 µm/nJ for latex and of 0.07 µm/nJ for iron. Yet, most of the knowledge of the mass and velocity distribution of dust, composition, flux and the effect of impacts in space instrumentation has been gathered in the last decades from dedicated experiments in space, such as dust detectors onboard GALILEO and ULYSSES, or the cosmic dust analyzer (CDA) aboard the CASSINI[123]. As an example, the CDA instrument has two sensors, the first one was a high-rate detector to count the number of particles and the second one analyzed the dust's charge, speed, size and direction. Another important source of knowledge of dust and debris characteristics has been obtained from satellites or parts thereof returned from space (Shuttle, Solar Max, Palapa, Westar, MIR, EURECA, HST). As an example, the chemical analysis of the craters on solar cells returned from the HST showed that caters with diameters of 100-3500 µm were produced by meteoroids, whereas craters with diameters of 1-100µm were produced mostly by space debris composed by aluminum and aluminum oxide, indicative of solid rocket motor fuel debris[125]. Special mention deserves the LDEF experiment, in which a tray with several optics was mounted in the exterior of the spacecraft (see Fig. 5a and b), then the optics were exposed 5 years and 8 months (32,422 orbits in LEO, from 842 km to 340 km) to micrometeoroids and space debris, and then the optics were returned and analyzed on Earth. This extended duration presented a unique opportunity to study the long-term effects of space exposure on the coatings and substrate materials flown. Among the most spectacular results, a 1-mm diameter impact in a bare $CaF_2$ substrate produced a 2-directions full cleavage, breaking the sample into 3 pieces, as shown in Fig. 5c. Another impact on a PbTe/ZnS multilayer-coated Ge substrate caused a coating delamination in the surroundings of the spallation area of 4.5 mm diameter; posterior analysis on the multilayer coating showed that the impact did not add stress or induce any further coating damage beyond the spallation area. The contamination of the SiO-coated Si substrate by aluminum provided the best example on LDEF of secondary ejecta and collateral effects of impacts. An impact occurred into the edge of the aluminum sample holder (see Fig. 5d), leaving secondary ejecta spray patterns of molten aluminum on the surface of the sample[126][127].

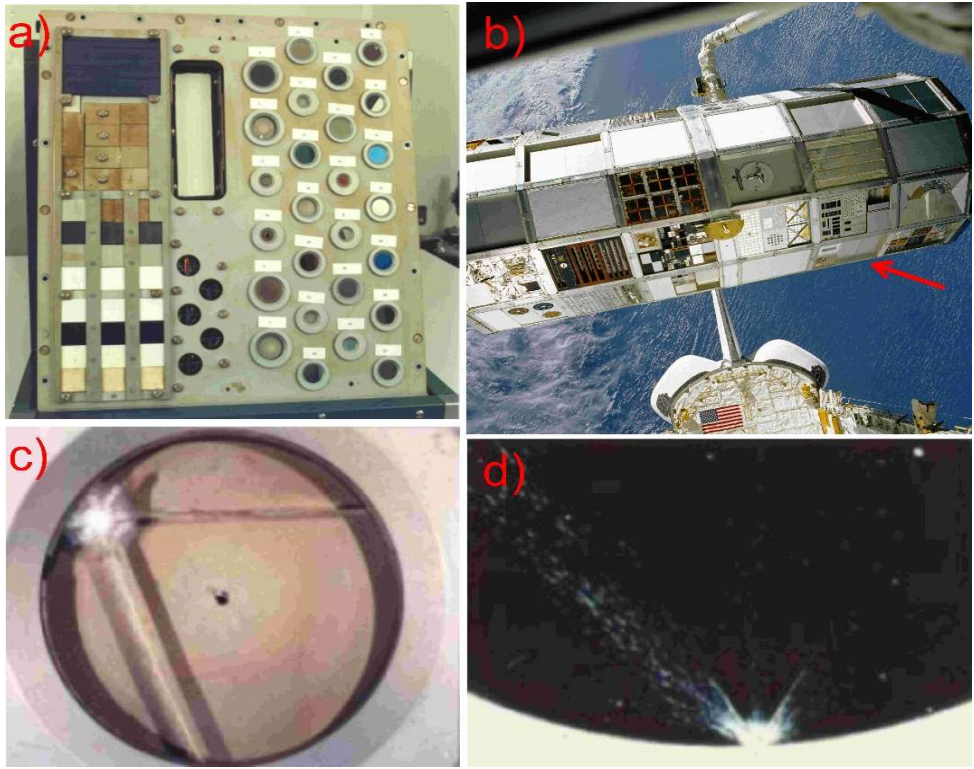

Figure 5: a) Tray B08 with several optics mounted. There were bare substrates and coated substrates, among other samples [https://www2.physics.ox.ac.uk, LDEF project]. b) The LDEF in orbit. The location of tray B08 in LDEF can barely be seen, but it is indicated with a red arrow [NASA Image and Video Library]. c) Impact and cleavage of the $BaF_2$ substrate [available at www.reading.ac.uk, Infrared Laboratory, LDEF]. d) Impact on the edge of the Al holder for the SiO-coated Si sample. Molten Al spray patterns can be seen on the sample surface.

Summarizing, space dust and debris can affect the performance of mirrors and coatings, or even produce a full mission failure. The body of spacecrafts can be protected from micrometeoroids and space debris impacts. F. Whipple proposed in 1947 [128] that a steel "skin" of one millimeter thickness spaced one inch from the main spacecraft body would disintegrate along with the high-velocity meteoroid upon impact, thus preventing the latter causing damage to the spacecraft. Even though this protective system has been verified, implemented, and improved since then, this cannot be used to protect the primary mirror of an optical telescope, for instance. Hence, for exposed optical systems protection relies mostly on prevention and prediction. It is common to all space agencies to pursue the common goal of reducing the generation of space debris from in-orbit explosions or collisions, discourage anti-satellite missile tests, and reduce the debris generated by rocket upper stages. In terms of prediction, models that can precisely account for (1) meteoroid velocity and mass distributions as a function of orbit altitude, (2) flux of meteoroids of larger sizes (>100 microns), (3) effects of plasma during impacts, and (4) variations in meteoroid bulk density with impact velocity, have been identified as a powerful tool to foresee the effect of dust and debris in future space observatories [119].

**3. Conclusions**

Optical coated elements for space instrumentation are mainly optimized in terms of efficiency and required working spectral band. After fabrication, witness samples undergo a series of laboratory tests required by the space agencies in order to qualify the components.

However, laboratory tests rarely reproduce the conditions in space, because the quality of the vacuum is not the same in space, some contaminants coming from the satellite itself are neglected, space environment is not always known precisely, the flux of particles and contaminants is lower in space than in accelerated tests. Moreover, very few experimental data are made available by the re-testing of components in those few cases in which the optics have been collected after a flight. The results of the qualification tests are rarely published and made available to the scientific community, as they are perceived as small technical details and because there is not a reference scientific journal which offers a solid background in this field, as opposite of the case of the electronics components.

On the other hand, both mirror substrates and coatings are a key component of space optics. Space mirrors must withstand a harsh environment, where servicing campaigns to clean or replace degraded optics are very limited or, most often, impossible. Mirrors must be able to keep acceptable performance through missions that may have a lifetime as long as decades. In fact, optical performance of the components strongly affects the scientific data outcomes, and their degradation can lead to a misinterpretation data due to an unknown change of the instrument radiometric response. In a more dramatic scenario, unpredicted mirror degradation may kill an expensive mission along with the strong expectations of the community for decades.

Hence, space opticians need to predict the behavior of coatings and substrates at the specific orbit and space conditions and for the full mission lifetime. In order to accomplish this, more experimental data need to be collected and shared. The clear definition of testing procedures to assess the robustness of optical components against the operational environment is pivotal to gain this knowledge, and thus for preventing in-flight failures, to fabricate robust coatings, or simply to model their degradation. In-situ testing experiments in which simple optical systems are coupled to the mirror optics for efficiency measurement over time during a flight could be an advantage. This paper is intended to contribute to build of a background knowledge in the field. Attention has been devoted to the main sources of mirror degradation: atomic oxygen, thermal processes, ultraviolet radiation, outgassing and cross-contamination, charged particles, and space debris and dust. An effort has been made to cite and comment the main literature on the degradation effects of all these sources on mirrors all over the electromagnetic spectrum, with emphasis at short wavelengths. Available information combines space simulation in the lab and also the heritage of six decades of space optics. Despite the long literature on space mirrors and degradation/stability issues, significant improvements are still desired for future space observatories. The development of large size and broadband mirrors will come together with new materials and coating designs. Future low-orbit to deep-space exploration will need to keep solving new issues on degradation resistance of mirrors.

**Author Contributions:** All authors have contributed to prepare the manuscript. They have read and agreed to the published version of the manuscript.

**Funding:** JIL acknowledges funding by Ministerio de Ciencia e Innovación, Gobierno de España (PID2019-105156GB-I00).

**Conflicts of Interest:** The authors declare no conflict of interest.